\documentclass[aps,pra,twocolumn,showpacs]{revtex4}
\usepackage{epsf}
\usepackage{amsmath}
\usepackage{bm}

\newcommand{\be}{\begin{equation}}
\newcommand{\ee}{\end{equation}} 
\newcommand{\ba}{\begin{eqnarray}}
\newcommand{\ea}{\end{eqnarray}}

\def\opone{\leavevmode\hbox{\small1\kern-3.8pt\normalsize1}}

\def\la{\langle}
\def\ra{\rangle}
\def\e{{\rm e}}
\def\i{{\rm i}}

\def\ga{\alpha}
\def\gl{\lambda}

\def\gd{\delta}
\def\gD{\Delta}
\def\go{\omega}
\def\gO{\Omega}
\def\d{\dagger}
\def\p{\partial}
\begin{document}

\title{Quantum statistics of overlapping modes in open resonators}

\author{Gregor Hackenbroich, Carlos Viviescas, and Fritz Haake}

\affiliation{Fachbereich Physik, Universit\"at Duisburg-Essen,
45117 Essen, Germany}

\date{\today}


\begin{abstract}
  We study the quantum dynamics of optical fields in weakly confining
  resonators with overlapping modes. Employing a recently developed
  quantization scheme involving a discrete set of resonator modes and
  continua of external modes we derive Langevin equations and a master
  equation for the resonator modes.  Langevin dynamics and the master
  equation are proved to be equivalent in the Markovian limit. Our
  open-resonator dynamics may be used as a starting point for a
  quantum theory of random lasers.
\end{abstract}

\maketitle

\section{Introduction} \label{sec.intro} The interaction of a single
resonator mode with an external optical field 
leads to damping and noise for the intracavity mode.  The textbook
example is a {\em single} oscillator linearly coupled to a continuum
of harmonic oscillators \cite{Gar00,Wal94,Wei99,Haa85}. Less understood
and in fact object of ongoing debate
\cite{Pet79,Bar88,Sie89,Gol91,Gra98,Bar99,Lam99,Lee00,Lam02} are the
damping and noise properties of {\em multimode} fields in resonators.
Multimode fields show an excess noise or Petermann factor
\cite{Pet79,Sie89,Bee98}. In laser systems this factor gives rise to a
peculiar enhancement of the laser linewidth above the fundamental
Schawlow--Townes value; this enhancement was measured in recent
experiments on unstable laser cavities \cite{Che96,Eij97}.  Excess noise
and the enhancement of the laser linewidth may be attributed to the
nonorthogonality and the spectral overlap of the cavity eigenmodes
\cite{Sie89} in the presence of the coupling to the external field.

While the phenomenon of excess noise has been known for more than 20
years, there is still no complete description of the quantum
statistics of overlapping modes. A few example systems were discussed
\cite{Gol91,Gra98} from a quantum mechanical point of view.  Recently
several authors \cite{Bar99,Lam99,Lee00,Lam02} proposed quantum Langevin
and master equations for multimode fields; however, the {\em status}
of these equations remained unclear as they were not derived from
rigorously quantized electromagnetic fields.  This is in contrast to
the quantum properties of a single-mode field (linearly coupled to an
external heat bath) which are known \cite{Gar00,Wei99,Haa85} for
arbitrary damping strength and arbitrary heat-bath temperature.

The goal of the present paper is to derive and clarify the status of
stochastic equations for the field dynamics in resonators with
overlapping modes.  The experimental motivation for our work derives
both from realizations \cite{Che96,Eij97} of unstable laser cavities
and from recent experiments of highly disordered dielectrics which
form mirrorless so-called random lasers \cite{Fro98,Cao99}. We address
the field dynamics both within the Heisenberg picture (in terms of
quantum Langevin equations) and within the Schr\"odinger picture
(employing a master equation for the reduced density matrix of the
cavity modes).  We go beyond previous work in the following respects:
(i) We derive the field dynamics starting from rigorously quantized
electromagnetic fields. In particular, no restriction of the
dimensionality and the vector character of the field strengths are
indulged in. Keeping only resonant terms in the field Hamiltonian and
adopting a Markov approximation for the memory of the ``bath'' (i.e.,
the electromagnetic field outside the resonator), we obtain the
equations of motion for the intracavity field, including expressions
for all damping and noise forces.  (ii) Our derivation clarifies the
status of the resulting stochastic equations. In particular, we show
that they correctly describe separated as well as spectrally
overlapping resonances. Corrections would only be important if the
resonance widths were to become comparable to the resonance
frequencies themselves, a regime not encountered in optical or even
microwave resonators. (iii) We derive a representation of our master
equation in terms of non-orthogonal modes and compare our result with
master equations proposed earlier \cite{Bar99,Lam99,Lam02}. This
allows us to specify the physical conditions under which those
previously proposed master equations hold.

\section{Field quantization for open-resonator geometries}
\label{sec.hamiltonian}

There is nothing to add to the familiar canonical quantization of the
electromagnetic field \cite{Gar00,Wal94}. The field components may be
expanded in terms of any complete set of basis functions. As a matter
of convenience one usually employs eigenmodes pertinent to the given
geometry and respecting the physical boundary conditions. Such
expansions are at issue here.

In the presence of a more or less open resonator, one often wants to
distinguish between ``inside'' and ``outside'' even though any
opening, i.e., a hole in the material walls of the resonator gives to
the latter concepts an element of arbitrariness. Any choice of a
fictitious surface covering the hole yields its own inside/outside
separation. Moreover, different boundary conditions may be imposed at
the chosen separating surface.  Nevertheless, each such surface and
boundary condition entail eigenmodes allowing us to represent the
electromagnetic field almost everywhere, with the qualifier ``almost''
reminding us of the fact that the expansion cannot be expected to
converge pointwise everywhere, and in particular not on an arbitrarily
chosen boundary.

The freedom in choosing the separating surface and the boundary
condition thereon may be used \cite{Sav02,Hac02,Viv03} to define a
discrete set of inside modes which vanish outside; and similarly a
continuum (or even several continua distinguished by a ``channel''
index) of outside modes which vanish inside and fulfill
scattering-type boundary conditions at infinity; a scattering
condition specifies a single channel for a wave coming in from
infinity and entails different amplitudes for the partial waves going
out through the various channels. Canonical quantization then amounts
to representing the coefficients of the mode expansion of the
electromagnetic field by creation and annihilation operators
$\{a_{\lambda}, a_\lambda^{\d} \}$ of photons of the $\lambda$-th
inside mode and likewise $\{b_m(\omega), b^\dagger_m(\omega)\}$ of
photons of the $(m,\omega)$th outside mode, where $m$ labels channels
(including polarization) and $\omega$ distinguishes modes within the
channel continuum. The bosonic commutators read
\begin{subequations} 
\label{2.1}
\begin{eqnarray}
[a_{\lambda}, a_{\lambda^\prime}^\dagger] &=& 
\delta_{\lambda,\lambda^\prime} \, ,  
\\
\label{2.1b}
\ [ b_{m}(\omega) , b_{m^\prime}^\dagger (\omega^\prime) ] &=& 
\delta_{mm'}\delta (\omega- \omega^\prime) \, ; 
\end{eqnarray}
\end{subequations}
moreover, inside operators commute with outside ones.

It was shown in two previous papers \cite{Hac02,Viv03} that the
Hamiltonian of the electromagnetic field can be rigorously expressed
as the following bilinear form in the foregoing creation and
annihilation operators,
\begin{eqnarray} H=&& \!\!\!\!\sum_{\gl}\hbar \go_\gl a_\gl^\d a_\gl
+\sum_{m}\int d\go\, \hbar\go\, b_m^\d (\go)b_m(\go) \nonumber \\
&& + \hbar  \sum_{\gl , m} \int d\go   [{\cal W}_{\gl m}
(\go)a_\gl^\d b_m(\go)  \nonumber \\
&&  +{\cal V}_{\gl m}(\go)a_\gl b_m(\go)+ {\rm H.c.}] . \label{2.2} 
\end{eqnarray} 
Here, the inside mode frequencies $\go_\gl$ as well as the
inside-outside coupling amplitudes ${\cal W}_{\gl m},{\cal V}_{\gl m}$
reflect the choices made for the separating surface. The latter
amplitudes are integrals, taken over the separating surface, of
products of inside (or outside) mode functions with derivatives of
outside (inside) mode functions. Specifically, if the von--Neumann
boundary condition
\begin{equation} \label{2.3}
{\bf n} \times \left[ \nabla \times {\bf u}_{\lambda} \right] 
\big|_{\partial I} = 0 
\end{equation}
was imposed for the inside modes ${\bf u}_\gl$ and the Dirichlet
boundary condition
\begin{equation} \label{2.4}
{\bf n} \times  {\bf v}_m(\omega) \big|_{\partial I} = 0 
\end{equation}
for the outside modes ${\bf v}_m(\go)$ at the boundary $\partial I$
(with normal vector ${\bf n}$), the coupling amplitudes are
given by
\begin{eqnarray}
\label{2.5}
\mathcal{W}_{\lambda m}(\omega) \!\!\!&=&\!\!\! \frac{c^2}{2
  \sqrt{\omega_\lambda \omega}} \int_{\partial
  I} \!\!\! d^2 r \, {\bf u}_\gl^*({\bf r}) \cdot \{ {\bf n} \times
  [\nabla \times {\bf v}_m(\go, {\bf r})]\} , \quad  \\
\label{2.6}
\mathcal{V}_{\lambda m}(\omega) \!\!\!&=&\!\!\! \frac{c^2}{2 
\sqrt{\omega_\lambda \omega}} \int_{\partial
  I} \!\!\! d^2 r \, {\bf u}_\gl({\bf r}) \cdot \{ {\bf n} \times
  [\nabla \times {\bf v}_m(\go, {\bf r})]\} . \quad
\end{eqnarray}
If the physical boundary conditions on material surfaces entail
time reversal invariance, the amplitudes ${\cal W,V}$ can be
chosen real and become identical, ${\cal W=V}$.

The derivation of the Hamiltonian takes full account of three spatial
dimensions as well as of the vector character of the electric and
magnetic fields ${\mathbf E}({\mathbf x},t)$ and ${\mathbf B}({\mathbf
  x},t)$; in brief, the Heisenberg equations of motion generated by
that $H$ are equivalent to the quantized Maxwell equations. Moreover,
by composing the inside (outside) mode functions with coefficients
$\{a_\gl,a_\gl^\d\}$ ($\{b_m(\go), b_m^\d (\go)\}$) one gets the fields
${\mathbf E}({\mathbf x},t)$ and ${\mathbf B}({\mathbf x},t)$ inside
(outside); truncated such compositions will be robust against small
changes of the separating surface unless an unreasonable choice was
made to begin with.
 
Hamiltonians of the above form have been used \cite{Bar88,Gar00} in
quantum optics for high-quality resonators; there, inside excitations
are but slowly dissipated to the outside, with life times long not
only compared to typical internal periods $2\pi/\go_\gl$ but even to
beat periods $2\pi/ \gD\go$, where $\gD\go$ denotes the modulus of the
typical frequency spacing of neighboring internal modes. In such
applications the antiresonant terms ${\cal V}_{\gl m}a_\gl b_m+{\cal
V}_{\gl m}^*a_\gl^\d b_m^\d$ can be neglected. However, the derivation
of Refs. \cite{Hac02,Viv03} also secures validity of Hamiltonian
(\ref{2.2}) for very much open resonators where the outside field
causes the inside resonances to overlap, the main case of interest in
the present paper.  The Hamiltonian (\ref{2.2}) even remains valid in
the extreme case of overdamped inside excitations, where mode
frequencies $\go_\gl$ are overwhelmed by large escape rates $\kappa$;
it is only in that extreme situation, which appears as not of interest
in optics, that the antiresonant terms would be important.

\section{Langevin equations for overlapping resonances}
\label{sec.langevin}

While Hamiltonian (\ref{2.2}) describes the coupled dynamics of the
inside and outside fields, its principle application is to separate
descriptions of those two subsystems. As already indicated above, the
continua of outside modes tend to act as a ``bath'' damping the
discrete inside modes. The effectively irreversible dynamics of the
inside modes becomes manifest when the bath degrees of freedom are
eliminated. The Heisenberg equations of motion of the inside
amplitudes $a_\gl$ , $a_\gl^\d$ then take the form of Langevin
equations, in which the outside amplitudes $b_m(\go)$, $b_m^\d (\go)$
enter only with their initial values in inhomogeneities, the Langevin
noise forces. Due to the bilinear form of Hamiltonian (\ref{2.2})
the Langevin equations can be determined rigorously by diagonalization
of $H$, i.e., without resorting to perturbation expansions (for a {\em
  single} ``system'' oscillator the diagonalization was performed in,
e.g., Refs. \cite{Ull66} and \cite{Haa85}).

The Langevin equations to be noted in the present section are
simplified in three respects. First, we bar all antiresonant terms,
assuming the absence of overdamping. Second, we confine the discussion
to a Markovian situation valid for times larger than ``bath
correlation times'' $\tau_{\rm bath}$ (like the thermal time
$\tau_{\rm bath}^{\rm th}=\hbar/k_BT$); the existence of a time scale
separation is thus assumed, such that all inside lifetimes are much
in excess of $\tau_{\rm bath}$. Third, to save space we do not bother
to pedantically write out the so--called frequency-shift terms which
are rarely needed in practice. The limit in question yields the
Langevin equations \cite{Viv03}
\begin{equation}
\label{3.1} \dot{a}_\gl(t)=-\i\go_\gl
a_\gl(t)-\sum_\mu\gamma_{\gl\mu}a_{\mu}(t) + F_\gl(t) \,, 
\end{equation}
wherein the damping matrix $\gamma_{\gl\mu}$ and the noise force
$F_\gl$ are given in terms of the inside-outside coupling amplitudes
${\cal W}_{\gl m}$ as
\begin{subequations} 
\label{3.2} 
\begin{eqnarray}
\gamma_{\gl\mu} &=& \pi \left({\cal W}{\cal W}^\d\right)_{\gl\mu}\, ,
\\ F_{\gl}(t) &=& -\i \int\! d\go \,\e^{-\i\go (t-t_0)}\sum_m {\cal
W}_{\gl m}\,b_m(\go,t_0)\,.
\end{eqnarray}
\end{subequations}
Note that the damping matrix $\gamma$ is non-negative and Hermitian;
the neglected frequency-shift terms would amount to an anti-Hermitian
addition to $\gamma$. Under conditions of time reversal invariance the
matrix $\gamma$ is even real symmetric. The Markovian limit mentioned
allows us to drop the frequency dependence of the coupling amplitudes
${\cal W}$ over the spectral range in consideration and to restrict
the time span passed since the initial moment $t_0$ as $t-t_0\gg
\tau_{\rm bath}$. In that limit the force has a white spectrum
according to $\la F_\gl(t)\ra=0$, and the noise
\begin{subequations} \label{3.3}
\begin{eqnarray}
\la
F_\gl^\d(t)F_{\mu}(t')\ra &= &2\gamma_{\mu\gl}\,n_{\rm
th}\,\delta(t-t'), \\
 \la F_\gl(t)F_{\mu}^\d(t')\ra &=& 2\gamma_{\gl\mu}(1+n_{\rm
th})\,\delta(t-t')\,.
\end{eqnarray}  
\end{subequations} 
The second two-time correlation functions follows from the first and
the commutation relations (\ref{2.1b}). The remaining second moments
vanish, $\la F_{\gl}(t)F_{\mu}(t') \ra=0=\la F_{\gl}^\d (t) F_{\mu}^\d
(t')\ra$, and higher-order moments follow from the ones of orders 1
and 2 according to Gaussian statistics. The thermal number of
photons $n_{\rm th}= [\exp(\hbar\bar{\go}/kT)-1]^{-1}$ appearing in
the second moments must be taken as frequency independent throughout
the spectral range of inside frequencies $\go_\gl$ under
consideration.  In the limit $k T \ll \hbar \bar{\omega}$ we recover
the Langevin equations of Bardroff and Stenholm \cite{Bar99}.

The appearance of a \textit{nondiagonal} damping matrix $\gamma
=\pi{\cal W}{\cal W}^\d$ signals that the Langevin equation may
legitimately be applied to the case of overlapping resonances, in
which typical matrix elements $\gamma_{\gl\mu}$ are larger than a
typical nearest-neighbor spacing $\Delta\go$ of frequencies. However,
all elements of $\gamma$ must be smaller in magnitude than the
frequencies $\go_\gl$ themselves as antiresonant terms were dropped in
the derivation of Eq.~(\ref{3.1}).  Conversely, we could specialize to
the weak-coupling regime $|\gamma_{\gl\mu}|\ll\Delta\go$. Then,
lowest-order perturbation theory simply amounts to dropping the
off-diagonal elements of the damping matrix, $\gamma_{\gl\mu}\to 0$
for $\gl\neq\mu$, whereupon the Langevin equation (\ref{3.1}) simplifies
so as to describe a set of mutually independent damped harmonic
oscillators.

For some applications it will be helpful to rewrite the above Langevin
equation with the non-Hermitian matrix
\begin{equation}
\label{3.4} {\cal H}_{\gl\mu}=\hbar \go_\gl\delta_{\gl\mu}-\i \hbar 
\gamma_{\gl\mu} 
\end{equation}
diagonalized. The eigenvalues of ${\cal H}$ will then represent
``true'' resonances of the cavity. As a ``penalty'' for that change of
representation one would have to work with non-standard commutation
relations for the operators connected with the eigenvectors of ${\cal
  H}$ (see Sec.\ \ref{sec.stat}).

A final remark on the dynamics of the outside field is in order here.
The time scale separation assumed in deriving the above Langevin
equations is effective outside as well as inside as the outside field
evolves in adiabatic equilibrium with the inside field for times
$t-t_0\gg\tau_{\rm bath}$. For the quantitative treatment by the
quantum optical input-output formalism or, equivalently, scattering
theory we refer the interested reader to \cite{Viv03,Gar00,Wal94}.

\section{Master equation for overlapping resonances}
\label{sec.master}

We now come to the central section of the present paper. Switching to
the Schr{\"o}dinger picture we present the master equation for the
reduced density operator $\rho(t)$ of the inside field equivalent to
the Langevin equation of the preceding section,
\begin{eqnarray} \label{4.1}
\dot{\rho} & &\!\!\!\!\!\!=\!\!-\i\sum_\gl\go_\gl [a_\gl^\d a_\gl,\rho] 
+(1+n_{\rm th}) \sum_{\gl\mu}\gamma_{\gl\mu}\left\{
[a_\mu,\rho a_\gl^\d] \right. \nonumber \\
&& \hspace*{-0.3cm}
\left. +[a_\mu\rho, a_\gl^\d]\right\} 
+n_{\rm th}\sum_{\gl\mu}\gamma_{\gl\mu}\left\{[a_\gl^\d,\rho a_\mu]
+[a_\gl^\d\rho, a_\mu]\right\} . \hspace*{0.7cm}  
\end{eqnarray} 
This equation generalizes the familiar quantum optical master equation
for a single damped harmonic oscillator to many oscillators coupled by
the (off-diagonal elements of the) damping matrix $\gamma_{\gl\mu}$.
The latter coupling is important when the damping is strong enough to
cause spectral overlap of modes. The first double sum, proportional to
$1+n_{\rm th}$, describes spontaneous and induced emission of photons
towards the outside while the second double sum, proportional only to
$n_{\rm th}$, describes absorption from the outside; that
interpretation is easily checked by employing the Fock representation,
i.e.\ the representation in terms of eigenstates of the photon number
operators $a_\gl^\d a_\gl$.

Systematic and stochastic forces are not as clearly separated here as
in the Langevin equation. In order to bring about such distinction
here as well, and to prepare for the proof of equivalence of
Eqs. (\ref{3.1}) and (\ref{4.1}) we may imagine the density operator
$\rho(t)$ at time $t$ antinormally ordered in the annihilation and
creation operators (all $a$'s to the left of all $a^\d$'s); further,
we rearrange the commutators in the right-hand side of the master
equation (\ref{4.1}) such that $\dot{\rho}$ becomes anti--normally
ordered provided $\rho(t)$ is,
\begin{eqnarray}
\label{4.2}
\dot{\rho}&=&-\i\sum_\gl\go_\gl\{ [a_\gl,\rho a_\gl^\d]-
                [a_\gl\rho, a_\gl^\d]\} \nonumber \\
        &&+\sum_{\gl\mu}\gamma_{\gl\mu}\{[a_\mu,\rho a_\gl^\d]
                  +[a_\mu\rho, a_\gl^\d]\} \nonumber \\
        &&+2n_{\rm th}\sum_{\gl\mu}\gamma_{\gl\mu}
               [[a_\mu,\rho], a_\gl^\d]\,.
\end{eqnarray}
The latter form of the master equation preserves antinormal ordering
of $\rho(t)$ at all times. Moreover, we have now separated reversible
drift terms (proportional to the frequencies $\go_\gl$), irreversible
drift terms ($\propto \gamma_{\gl\mu}$) not involving the thermal
number of quanta $n_{\rm th}$, and noise generated diffusion terms
($\propto n_{\rm th}$); the latter interpretation will become obvious
in the following section where a representation based on coherent
states will be employed.

\section{Equivalence of Langevin and master equations}
\label{sec.equivalence}

There are various ways of demonstrating the equivalence of the
Langevin equations (\ref{3.1}) and the master equation (\ref{4.1}) and
(\ref{4.2}).  A somewhat indirect (and, in fact, the most laborious)
way would be to start from the Liouville--von Neumann equation
$\dot{\rho}_{{\rm in}\oplus {\rm out}}(t)=-\i[H,\rho_{{\rm in} \oplus
{\rm out}}(t)]/\hbar$ for the density operator of the full
electromagnetic field and stick to the Schr\"odinger picture in
eliminating the outside field, employing the same approximations as in
deriving the Langevin equations.  A standard, and more economical,
procedure is to show that Eq.~(\ref{3.1}) and (\ref{4.2}) entail the same
evolution equations for all mean values
$\la\prod_\gl(a_\gl^\d)^{m_{\gl}}\prod_\mu (a_\mu)^{n_\mu}\ra(t)$ with
integer exponents $\{m_\gl,n_\mu\}$. We here prefer to employ a less
familiar but particularly elegant method.

If we consistently stick to antinormal ordering of $\rho(t)$ we may
write the commutators in the master equation (\ref{4.2}) as differential
operators as $[a,(\cdot)]\to (\p /\p a^\d)(\cdot)$ and
$[(\cdot),a^\d]\to (\p/\p a)(\cdot)$. We may then just as well
degrade all creation and annihilation operators to complex c-number
variables and the density operator to a real function of those
variables,
\begin{equation} \label{5.1} a_\gl\to\ga_\gl\,,\quad
a_\gl^\d\to\ga_\gl^*\,,\quad \rho(t)\to P(\{\ga,\ga^*\},t)\,.
\end{equation} 
The function $P$ in question is the familiar Glauber--Sudarshan
quasiprobability \cite{Gar00} which has as its moments expectation values
of normally ordered observables,
\begin{eqnarray} \label{5.2}
&&\Big\la\prod_\gl(a_\gl^\d)^{m_{\gl}}\prod_\mu (a_\mu)^{n_\mu}\Big\ra(t)= \\
\nonumber 
&& \int \left\{\prod_\gl d\ga_\gl d\ga_\gl^*(\ga_\gl^*)^{m_{\gl}}
(\ga_\gl)^{n_\gl} \right\} P(\{\ga,\ga^*\},t).  
\end{eqnarray} 
The master equation (\ref{4.2}) then becomes the Fokker--Planck equation
(with the shorthand $\p/\p \ga_\gl\equiv \p_\gl$, $\p/\p \ga_\gl^*\equiv
\p_\gl^*$)
\begin{eqnarray} \label{5.3}
\dot{P}=\Big[\!\!\!\!\!&& -\i\sum_\gl\go_\gl(\p_\gl^*\ga_\gl^*-\p_\gl\ga_\gl)
 \\
&&+\sum_{\gl\mu}\gamma_{\gl\mu}\big(\p_\gl^*\ga_\mu^*+\p_\mu\ga_\gl
+2n_{\rm th}\p_\mu\p_{\gl}^*\big)\Big]P\,. \nonumber  
\end{eqnarray}
At this point the interpretation of the various terms in the master
equation (\ref{4.2}) given earlier becomes obvious.

In the same vein we may degrade the Langevin equation to the c-number
equation
\begin{equation} \label{5.4} \dot{\ga}_\gl(t)=-\frac{\i}{\hbar}\sum_\mu {\cal
H}_{\gl\mu}\ga_\mu(t)+\varphi_\gl(t), 
\end{equation} 
where $\varphi_\gl(t)$ is the c-number representative of the operator
valued random force $F_\gl(t)$. Gaussian statistics and zero mean are
imparted to $\varphi(t)$. Moreover, since observables are consistently
to be taken as normally ordered we must identify second moments as
\begin{equation}
\label{5.5} \la\varphi_\gl^*(t)\varphi_\mu(t')\ra=\la
F_\gl^\d(t)F_\mu(t')\ra= 2n_{\rm th}\gamma_{\mu\gl}\delta(t-t')\,;
\end{equation}
while holding on to $\la\varphi_\gl(t)\ra=0,\, \la\varphi_\gl(t)
\varphi_\mu(t')\ra=0$ and Gaussian factorization for the
higher-order moments.

At this point the equivalence proof can be carried out by deducing the
Fokker--Planck equation (\ref{5.3}) from the c-number Langevin equation
(\ref{5.4}), and to that task we now turn. In order to make the argument
as transparent as possible we drop mode indices; the initial time at
which statistical independence of the inside and outside fields is
assumed is taken as $t_0=0$. We start by considering a fixed
realization of the time-dependent ``external forces''
$\varphi(t),\varphi^*(t)$. The solution of the Langevin equation is
then a functional of the temporal course taken by the external force
between the initial and the current time.  Correspondingly, we
introduce a density $W(\ga,\ga^*,t|\{\varphi(t),\varphi^*(t)\})$
functionally depending on the realization of the external field under
consideration; at the initial time $t=0$ the new density coincides
with the Glauber--Sudarshan $P$-function, $W(0)=P(0)$.  The density
$W$ obeys the evolution equation
\begin{eqnarray} \label{5.6} \dot{W}(t)&=&\big(L_0
+l(t)\big)W(t) \, ,\\ \nonumber L_0
&=&-\i\go(\p^*\ga^*-\p\ga)+\gamma(\p^*\ga^*+\p\ga) \, ,\\ \nonumber l(t)
&=& -\big(\p^*\varphi^*(t)+\p\varphi(t)\big)\,.  
\end{eqnarray} 
Clearly, we encounter a first-order partial differential equation
since the random character of the external force is not yet invoked.
The time independent part $L_0$ of the generator accounts for
oscillation and damping, while $l(t)$ includes the time dependent
forces $\varphi(t),\varphi^*(t)$. The formal solution reads $W(t)=
\big(\exp{\int_0^tds(L_0+l(s))}\big)_+W(0)$, where $(\cdot)_+$ denotes
positive time ordering. Temporarily employing an ``interaction
picture'' we split off the motion generated by $L_0$ as
\begin{eqnarray} \nonumber
W(t)&=&\e^{L_0t}\tilde{W}(t)\,,\\ \label{5.7}
\tilde{W}(t)&=&\Big[\exp{\int_0^t\!ds\,\tilde{l}(s)}\Big]_{\!+}P(0)\,,\\
\nonumber
\tilde{l}(t)&=&\e^{-L_0t}l(t)\e^{L_0t}=-\big[\varphi^*(t)\tilde{\p}^*(t)
+\varphi(t)\tilde{\p}(t)\big]\,.  
\end{eqnarray} 
Now we may take the average over the Gaussian ensemble of realizations
of $\varphi(t),\varphi^*(t)$. The average of the $\varphi$-dependent
density $W(t)$ is just the Glauber--Sudarshan $P$-function, $\la
W(\ga,\ga^*,t|\{\varphi(t), \varphi^*(t)\})\ra=P(\ga,\ga^*,t)$. Invoking
the above first and second moments we get
\begin{eqnarray} \nonumber
\tilde{P}(t)&\!\!=&\!\!\!\!\Big\la\left(\exp\left[{\int_0^t\!ds\tilde{l}(s)}\right]
\right)_{\!+}\Big\ra P(0)\\ 
\label{5.8} &=&\!\!\!\!\left(\exp\left[{\int_0^t\!\!ds\int_0^t\!\!\!ds' \la\varphi^*(s)
\varphi(s')\ra\tilde{\p}^*(s)\tilde{\p}(s')}\!\right]\!\right)_{\!+}P(0) 
\nonumber \\
&=&\!\!\!\!\left(\exp\left[{2n_{\rm th}\gamma\int_0^t\!ds
\tilde{\p}^*(s)\tilde{\p}(s) }\right] \right)_{\!+}P(0)\,.
\end{eqnarray}
Upon differentiating with respect to time we have $\dot{\tilde{P}}
(t)=2n_{\rm th}\gamma \tilde{\p}^*(t) \tilde{\p} (t)\tilde{P}(t)$, and
going back to $P(t)=\e^{L_0t}\tilde{P}(t)$, we arrive at the
Fokker--Planck equation (\ref{5.3}) and have thus arrived at our goal.

We should add that a mathematically more satisfactory variant of the
foregoing considerations would result from employing the so-called
Ito calculus \cite{Gar00}.

\section{Stationary solution of the master equation}
\label{sec.stat} The general time-dependent solution of the Fokker--Planck
equation (\ref{4.2}) is, due to the linearity of the drift coefficients
and the constancy of the diffusion matrix, easy to construct
\cite{Ris89}. We are in fact facing a stochastic process of the
Ornstein--Uhlenbeck type. The Gaussian distribution of the noise
together with the linear evolution equation (\ref{5.4}) imply that the
stochastic variables must be Gaussian distributed. In particular, the
stationary $P$ function is immediately checked to be
\begin{equation} \label{6.1} 
\bar{P} (\{ \alpha, \alpha^*\}) = \prod_{\lambda}\frac{1}{\pi n_{\rm
th}} \exp( -\alpha^*_\lambda \alpha_\lambda /n_{\rm th}) .
\end{equation} 
The dissipative coupling of the system modes is no longer visible in
the stationary state; rather, we encounter the thermal equilibrium
state one would also find in the absence of spectral overlap.

\section{Status of Langevin and Master equations and 
  related literature} \label{sec.status} Multimode fields are more complex
than single-mode fields due the additional frequency scale set by the
mean frequency spacing $\Delta \omega$ of internal modes. For
resonators with overlapping modes the mode spectral broadening is
comparable to $\Delta \omega$.  Langevin or master equations may then
only be used if they provide a {\em nonperturbative} description of
damping and noise (in the sense that the mode decay rates $\kappa$ may
exceed the mode spacing $\Delta \omega$).

To show that the Langevin (\ref{3.1}) and the master equations
(\ref{4.1}) and (\ref{4.2}) are non--perturbative in the above--mentioned
sense, we now drop the rotating--wave and Markov approximation and
address the exact field dynamics.  From the field Hamiltonian
(\ref{2.2}) one obtains the following set of Langevin equations
\begin{eqnarray}
\label{7.1} \dot{a}_\gl(t) &=& -\i\go_\gl
a_\gl(t)-\sum_\mu \int_{t_0}^t d t^\prime \left[
\Gamma_{\gl\mu} (t-t^\prime)
a_\mu (t^\prime) \right. \nonumber \\
&& \left. + \Sigma_{\gl\mu}(t-t^\prime) a_\mu^\dagger (t^\prime) 
\right] + f_\lambda(t).
\end{eqnarray}
Non--Markovian effects and corrections to isolated--mode behavior
are encoded in the autocorrelation functions
\begin{eqnarray}
\label{7.2}
\Gamma_{\gl\mu} (t-t^\prime) &=& \sum_m \int_{\omega_m}^\infty d
\omega \left[ \e^{-\i \omega (t-t^\prime)} W_{\gl m} (\omega) W_{\mu
m}^* (\omega) \right. \nonumber  \\
&& \left. \hspace*{0.7cm} - \e^{+\i \omega (t-t^\prime)} V_{\gl m}^* 
(\omega) V_{\mu m} (\omega) \right] , \\
\label{7.3}
\Sigma_{\gl\mu} (t-t^\prime) &=& \sum_m \int_{\omega_m}^\infty d
\omega \left[ \e^{-\i \omega (t-t^\prime)} W_{\gl m} (\omega) V_{\mu
m}^* (\omega) \right. \nonumber \\
&& \left. \hspace*{0.7cm} - \e^{+\i \omega (t-t^\prime)} V_{\gl m}^* 
(\omega) W_{\mu m}^* (\omega) \right] .
\end{eqnarray}
The noise term takes the form
\begin{eqnarray}
\label{7.4}
f_\lambda(t) &=& -\i \sum_m \int_{\omega_m}^\infty d \omega \left[ 
\e^{-i \omega (t-t_0)} W_{\gl m} (\omega) b_m(\omega, t_0) \right. 
\nonumber \\
&& \left. \hspace*{0.7cm} + \e^{+\i \omega (t-t_0)} V_{\gl m}^* 
(\omega) b_m^\dagger (\omega, t_0) \right].
\end{eqnarray}
Due to the bilinear form of the field Hamiltonian the exact Langevin
equations are linear in the system and bath operators. In principle,
their solution can be found by Laplace transformation.  The details of
the exact solution will depend both on the spectral strength of the
bath and the frequency dependent coupling amplitudes $W_{\gl
m}(\omega)$ and $V_{\gl m}(\omega)$.

The result (\ref{7.1})--(\ref{7.4}) holds for arbitrary damping. Here, our
main focus is on quantum optical systems where the oscillation
frequencies $\bar{\omega}$ much exceed the typical damping rates
$\kappa$. Then our result can be used to compute corrections to the
Markovian dynamics of Secs.\ \ref{sec.langevin}--\ref{sec.equivalence}. We
substitute the ansatz $a_\gl (t) \sim \exp(-\i [\bar{\go} -i \kappa]
t)$ with the damping rate $\kappa$ into Eq.\ (\ref{7.1}); here
$\bar{\go}$ includes a frequency shift $\delta \bar{\go}$. One obtains
four terms proportional to $a_\mu$ or $a_\mu^\dagger$, out of which
three terms are strongly oscillating: the contribution $\propto V^* W
\, (W V^*)$ has the oscillatory integrand $\sim \e^{- 2 \i \bar{\go}
t^\prime} \, (\e^{+ 2 \i \bar{\go} t^\prime})$, respectively, and the
term $\propto V^* W$ oscillates $\sim \e^{+ 2 \i \bar{\go} t}$ with
respect to $a_\lambda(t)$. To leading order in the ratio $| \delta
\gO|/\bar{\go} \equiv |\delta \bar{\go}-\i \kappa|/\bar{\go}$ only the
term $\propto W W^*$ must be kept. The corrections of higher order in
$|\delta \gO|/\bar{\omega}$ are very small for the systems of interest
in quantum optics. We note that there are no corrections of order
$|\delta \gO|/\Delta \go$; thus our field dynamics correctly describes
the regime of overlapping modes.

So far we described the field dynamics in terms of eigenmodes of a
conveniently chosen {\em closed} system. The associated mode operators
$\{ a_\gl, a^\d_\gl\}$ obey canonical commutation relations. The
master equation (\ref{4.1}) may equivalently be expressed in terms of a
biorthogonal set of modes representing the resonances of the {\em
open} cavity. It turns out that the operators associated with the
resonances obey nonstandard commutation rules. We focus on the case
$k T \ll \hbar \bar{\omega}$ and write Eq.~(\ref{4.1}) in the form
\begin{eqnarray} \label{7.5}
\dot{\rho}&=&-\frac{\i}{\hbar}(H_{\rm eff} \rho - \rho 
H_{\rm eff}^\dagger) + 2 \sum_{\gl\mu} \gamma_{\gl\mu}
a_\mu \rho a_\gl^\d \, ,\\
H_{\rm eff} &=& \sum_{\gl\mu} {\cal H}_{\gl\mu} a_\gl^\d 
a_\mu \,. 
\end{eqnarray} 
The non-Hermitian matrix ${\cal H}$ was introduced in Eq.\ (\ref{3.4}).
The complex eigenvalues of ${\cal H}$ represent the resonances of
cavity in the presence of the coupling to the bath.  Generally, the
eigenvalues are nondegenerate, and ${\cal H}$ can be diagonalized by
a similarity transformation ${\cal H}=T \Omega T^{-1}$. The diagonal
matrix $\Omega$ comprises the eigenvalues of ${\cal H}$ while $T$ is a
general matrix with complex entries.  In terms of two novel sets of
operators
\begin{eqnarray} \label{7.6}
d_n&=&\sum_\gl T^{-1}_{n \gl} a_\gl \, ,\\
\label{7.7}
e_n^\d&=&\sum_\gl a_\gl^\d T_{\gl n} \, ,
\end{eqnarray} 
the effective Hamiltonian takes the diagonal form
\begin{equation} \label{7.8}
H_{\rm eff} = \sum_n \Omega_n e_n^\d d_n \,.
\end{equation}
This result resembles the diagonal field Hamiltonian of closed
resonators. However, the peculiar properties of the open resonator
dynamics are encoded in nonstandard commutation relations of the new
operators: From Eqs.\ (\ref{7.6}) and (\ref{7.7}) and the canonical
commutation rules for the operators $\{ a_\gl, a^\d_\gl\}$ one finds,
e.g., $[ d_m, e_n^\d ] = \gd_{m,n}$ and $[ e_m, e_n^\d ] = A_{mn}$
with $A \equiv T^\d T$.  Substitution of Eq.\ (\ref{7.6}) into Eq.\ 
(\ref{7.5}) yields the master equation
$$
\dot{\rho}=-\frac{\i}{\hbar} ( H_{\rm eff} \rho - \rho 
H_{\rm eff}^\dagger) + \frac{\i}{\hbar}\sum_{nm} A_{nm} 
(\Omega_m-\Omega_n^*)  d_m \rho d_n^\d \,.
$$ 
This is the master equation proposed by Lamprecht and Ritsch
\cite{Lam02}. Our derivation provides a microscopic
basis for that equation \cite{remark} and proves the equivalence to the
master equation (\ref{4.1}) in the limit $k T \ll \hbar \bar{\go}$.

\begin{acknowledgments}
  We thank H.--J.\ Sommers and D.\ V.\ Savin for helpful discussions.
  This work has been supported by the SFB/TR 12 der Deutschen
  Forschungsgemeinschaft.
\end{acknowledgments}



\begin{thebibliography}{10}
\bibitem{remark} Our derivation shows that the master equation
  proposed in Ref.~\cite{Lam99} may describe the field dynamics only in the
  case of complete mode degeneracy.
\end{thebibliography}
\end{document}